\documentclass[aps,prl,amsmath,amssymb,twocolumn,floatfix]{revtex4-1}
\usepackage{graphicx}% Include figure files
\usepackage{dcolumn}% Align table columns on decimal point
\usepackage{bm}% bold math

\begin{document}

%%%%%%%%%%%%%%%%%%%%%%%%%%%%%%%%%%%%%%%%%%%%%%%%%%%
\title{Quasi-Static Internal Magnetic Field Detected in the Pseudogap Phase of Bi$_{2+x}$Sr$_{2-x}$CaCu$_2$O$_{8+\delta}$ by $\mu$SR}

\author{A.~Pal,$^1$ S. R.~Dunsiger,$^1$ K.~Akintola,$^1$ A. C. Y.~Fang,$^1$ A.~Elhosary,$^1$ M.~Ishikado,$^2$ H.~Eisaki,$^3$ and J. E.~Sonier,$^{1,4}$}

\affiliation{$^1$Department of Physics, Simon Fraser University, Burnaby, British Columbia, Canada V5A 1S6 \\
$^2$Neutron Science and Technology Center, Comprehensive Research Organization for Science and Society (CROSS), Tokai, Naka, Ibaraki, Japan 319-1106 \\
$^3$National Institute of Advanced Industrial Science and Technology, Tsukuba, Ibaraki, Japan 305-8568 \\
$^4$Canadian Institute for Advanced Research (CIFAR), Toronto, Ontario, Canada M5G 1Z8}

\date{\today}
%%%%%%%%%%%%%%%%%%%%%%%%%%%%%%%%%%%%%%%%%%%%%%%%%%%%%%%
\begin{abstract}
We report muon spin relaxation ($\mu$SR) measurements of optimally-doped and overdoped Bi$_{2+x}$Sr$_{2-x}$CaCu$_2$O$_{8+\delta}$ (Bi2212)
single crystals that reveal the presence of a weak temperature-dependent quasi-static internal magnetic field of electronic origin
in the superconducting (SC) and pseudogap (PG) phases. In both samples the internal magnetic field persists up to 160~K, but muon diffusion prevents
following the evolution of the field to higher temperatures. We consider the evidence from our measurments in support 
of PG order parameter candidates, namely, electronic loop currents and magnetoelectric quadrupoles.       
\end{abstract}

\maketitle
%%%%%%%%%%%%%%%%%%%%%%%%%%%%%%%%%%%%%%%%%%%%%%%
The origin of the PG in high transition temperature ($T_c$) cuprate superconductors is an enduring mystery, but is
widely believed to be a manifestation of a thermodynamic transition to a phase that breaks various symmetries.
The apparent detection of an ubiquitous IUC magnetic order in various cuprate families
by polarized neutron diffraction \cite{Fauque:06,Mook:08,Li:08,Baledent:10,Baledent:11,Li:11,Didry:12,Thro:14,Thro:15,Thro:17} lends support
to a longstanding idea that electronic loop-current order with translation symmetry forms in the PG phase \cite{Varma:97,Varma:06,Varma:14}. 
The existence of a magnetic order in the PG phase is supported by the observation of the Kerr effect in YBa$_2$Cu$_3$O$_y$ (Y123) \cite{Xia:08,Kapitulnik:09}.
Motivated by these results and other loop-current order predictions \cite{Chakravarty:01},
there have been numerous unsuccessful attempts to detect intrinsic magnetic order in the PG phase of high-$T_c$ cuprates 
by $\mu$SR \cite{Sonier:01,Sonier:02,MacDougall:08,Sonier:09,Huang:12,Pal:16,Zhang:17}. 
Such investigations of Y123 must account for the known affects of
charge-density-wave (CDW) order in the CuO chains, a potential unbuckling of the CuO$_2$ layers, and 
muon diffusion \cite{Sonier:02,Nishida:90,Sonier:17}, as well as magnetic correlations due to oxygen vacancies
in the CuO chains \cite{Yamani:06}. In hindsight Y123 is nonideal, because a weak magnetic contribution to the $\mu$SR signal
associated with the PG state cannot be disentangled from the influence of these other phenomena.
Although $\mu$SR studies of La$_{2-x}$Sr$_x$CuO$_4$ (La214) \cite{MacDougall:08,Huang:12} are limited only by muon diffusion at
high temperatures, the failure to detect PG magnetic order is not surprising given that
only short-range, two-dimensional IUC magnetic order is observed by neutrons in an $x \! = \! 0.085$ sample 
far below $T^*$ \cite{Baledent:10}. 

At odds with the above neutron studies of the PG phase is the recent failure to detect magnetic order 
in high-quality underdoped Y123 single crystals by polarized neutron diffraction \cite{Croft:17}.   
Moreover, it has been proposed that the above neutron experiments are not directly detecting magnetic dipolar ordering,
but rather quasi-static ordering of magnetoelectric multipoles \cite{Lovesey:15}. In this scenario, internal magnetic 
fields generated by the magnetoelectric quadrupole ordering are estimated to be below the detection limit of
nuclear magnetic resonance (NMR), but potentially detectable by $\mu$SR \cite{Fechner:16}. 

We have carried out zero field (ZF) and longitudinal field (LF) $\mu$SR measurements on a more promising system, namely, 
optimally-doped and overdoped Bi2212 single crystals with $T_c \! = \! 91$~K (OP91) 
and $T_c \! = \! 80$~K (OD80), respectively. The Bi2212 single crystals were grown using the 
traveling-solvent-floating-zone method \cite{Hobou:09}. The doping level was adjusted by tuning the excess oxygen content. 
Optimal doping was attained by annealing the single crystals in air at 720~$^{\circ}$C for 68 hours, and overdoping 
by annealing in an oxygen partial pressure of 2.3~atm for 72-250 hours at 400~$^{\circ}$C. 
The $T_c$ values were determined by measurements using a SC quantum interference device magnetometer. 
The OP91 sample consisted of seven single crystals of total mass 35 mg, double-stacked into a mosaic of total area 
$\sim \! 25$~mm$^2$. To ensure the muons did not pass through the sample, 0.025~mm thick high-purity silver (Ag) foil was 
placed in front of the sample to degrade the muon momentum. The OD80 sample consisted of three single crystals of total mass 72 mg,
two of which were double stacked, resulting in a total sample area of $\sim \! 40$~mm$^2$. 
The OD80 sample was thick enough to negate the need for Ag degrader foil. 
All $\mu$SR asymmetry spectra were recorded with the initial muon spin polarization {\bf P}(0) (and LF) parallel to the $\hat{c}$-axis.
The spectra are of the form $A(t) \! = \! a_0 G_z(t)$, where $a_0$ is the initial asymmetry and $G_z(t)$ is a relaxation function describing 
the time evolution of the muon spin polarization by the local magnetic fields sensed by the implanted muon ensemble.
   
\begin{figure}
\centering
\includegraphics[width=10.0cm]{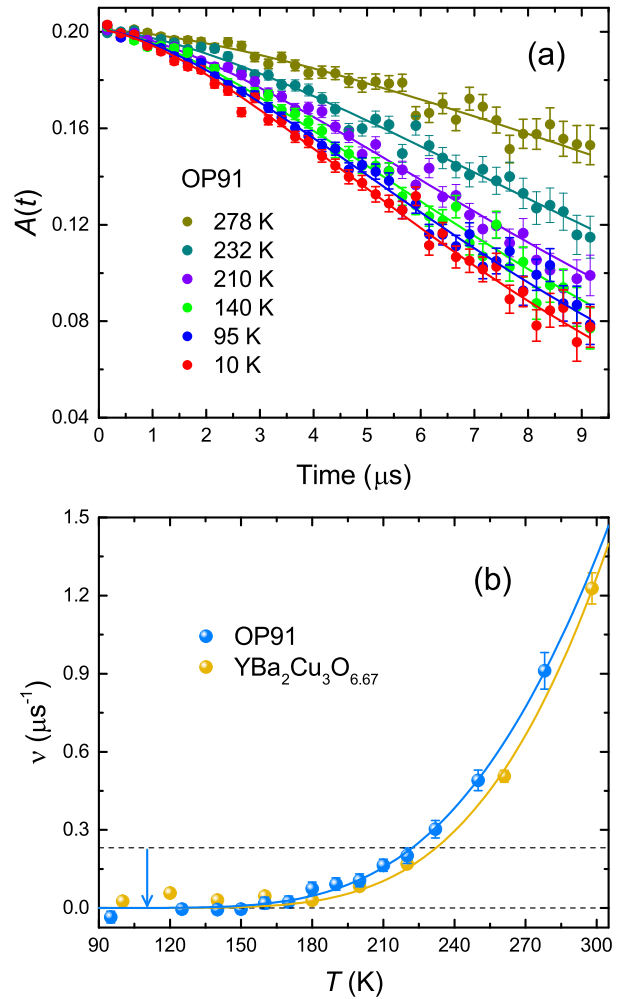}
\caption{(Color online) (a) Representative ZF-$\mu$SR asymmetry spectra recorded on the OP91 sample at different temperatures. 
The solid curves are fits to Eq.~(\ref{eqn:GKTexp}). (b) Temperature dependence of the muon hop
rate $\nu$ in the OP91 sample, and similar data for underdoped Y123 from Ref.~\cite{Sonier:02}. For the purpose of
comparison, the data for the OP91 sample are shifted downward by 0.232~$\mu$s$^{-1}$. The solid curves are fits to an Arrhenius
equation $\nu \! = \! A \exp(-E_0/k_{\rm B}T)$, which yield an activation energy of $E_0 \! = \! 130(4)$~meV for the OP91 sample and 
$E_0 \! = \! 151(9)$~meV for underdoped Y123.}
\label{fig1}
\end{figure}

Figure~\ref{fig1}(a) shows representative ZF-$\mu$SR asymmetry spectra recorded for the OP91 sample. The significant decrease
in the relaxation rate at the highest temperatures is due to muon diffusion. To show this is the case, the spectra at 
high $T$ were fit to the strong-collision dynamic Gaussian Kubo-Toyabe relaxation function $G_{\rm KT}(\Delta, \nu, t)$, 
where $\Delta$ is the static linewidth of a Gaussian field distribution and $\nu$ is the muon hop rate, or equivalently the average rate 
at which there are changes in the local magnetic field sensed by the muon \cite{Hayano:79}. 
The temperature dependence of the fitted muon hop rate $\nu$ with a zero offset correction is shown in Fig.~\ref{fig1}(b), 
together with earlier results for underdoped Y123 \cite{Sonier:02}. 
The zero offset of $\nu$ suggests the magnetic field distribution in Bi2212 is not as close to a Gaussian as in Y123.
Such deviations from a Gaussian form are to be anticipated, have been observed in La214 \cite{Huang:12}, and can be quantitatively
accounted for if the muon site(s) and electric field gradients in the material are known. With the zero offset correction,
the hop rate in Bi2212 obeys an Arrhenius relationship similar to Y123. In both compounds $\nu$ increases above 
$T \! \sim \! 160$~K, indicating a similar onset temperature for muon diffusion. In cuprates the muon is known to
form an O-H like bond with an oxygen ion. The results here imply that the thermal energy required to break the O-$\mu$ bond in Bi2212 is 
comparable to Y123. Unfortunately, muon diffusion masks clear evidence of PG magnetic order above 160~K. 

\begin{figure}
\centering
\includegraphics[width=10.0cm]{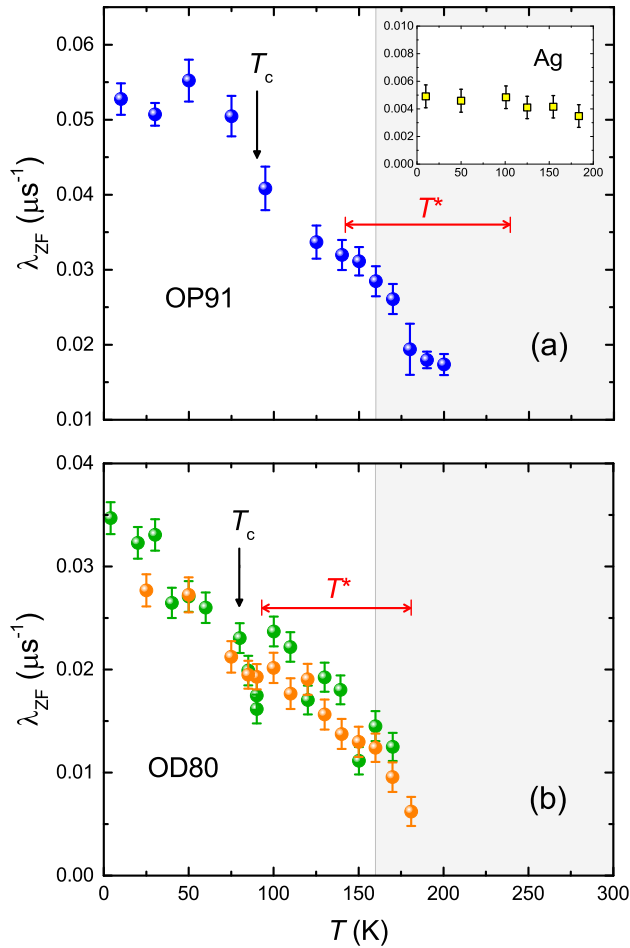}
\caption{(Color online) Temperature dependence of the exponential ZF relaxation rate in the (a) OP91 and (b) OD80 samples. 
The two data sets for the OD80 sample are from measurements performed during different beam periods. The range of values
for the PG temperature $T^*$ are from Ref.~\cite{Vishik:12}. The inset in (a) shows the ZF relaxation rate measured separately
in a $5.45 \! \times \! 6.54 \! \times \! 0.25$~mm sample of 99.998~\% pure Ag foil. The values of $\lambda_{\rm ZF}$ for Ag come
from fits to a single exponential relaxation function.}
\label{fig2}
\end{figure}

The ZF-$\mu$SR asymmetry spectra are well described by the product of an exponential relaxation function and a
static Gaussian ZF Kubo-Toyabe function intended to account for the nuclear-dipolar contribution. Specifically,
\begin{equation}
G_z(t)= G_{\rm KT}(\Delta, t) \exp(-\lambda_{\rm ZF} t) \, .
\label{eqn:GKTexp}
\end{equation}
Global fits of the ZF-$\mu$SR spectra as a function of temperature with $\Delta$ as a common parameter yield
$\Delta \! = \! 0.0958(6)$~$\mu$s$^{-1}$ and $\Delta \! = \! 0.095(1)$~$\mu$s$^{-1}$ for the OD80 and OP91 samples, respectively.
Figure~\ref{fig2} shows the temperature dependence of $\lambda_{\rm ZF}$.
For comparison, the inset of Fig.~\ref{fig2}(a) shows
$\lambda_{\rm ZF}$ versus $T$ measured in a sample of 99.998~\% pure Ag comparable in size to the Bi2212 samples.
The ZF relaxation rate in Ag is solely due to weak nuclear dipole fields, and is essentially negligible.
However, a $T$-dependent relaxation rate can arise from thermal contraction of the sample holder, which slightly changes the position 
of the sample in the three orthogonal pairs of Helmholtz coils used to cancel the external magnetic field. The results on 
Ag show that any such change in $\lambda_{\rm ZF}$ between 184~K and 10~K is less than $0.0014$~$\mu$s$^{-1}$.     
On the other hand, $\lambda_{\rm ZF}$ in the Bi2212 samples exhibit a significant increase with decreasing $T$ below 160 K.
Such behavior reflects a change in the linewidth of the internal magnetic field distribution sensed by the muon ensemble,
which {\it cannot} be explained in terms of muon diffusion alone. This is our main finding.

The values of $T^*$ for Bi2212 are ill defined. The temperature ranges for $T^*$ indicated in
Fig.~\ref{fig2} come from a compilation of values measured by different techniques \cite{Vishik:12}. Because the range of 
experimental values for $T^*$ extend above 160~K, we cannot say whether there is a spontaneous ZF relaxation appearing at the 
PG onset.

\begin{figure}
\centering
\includegraphics[width=10.0cm]{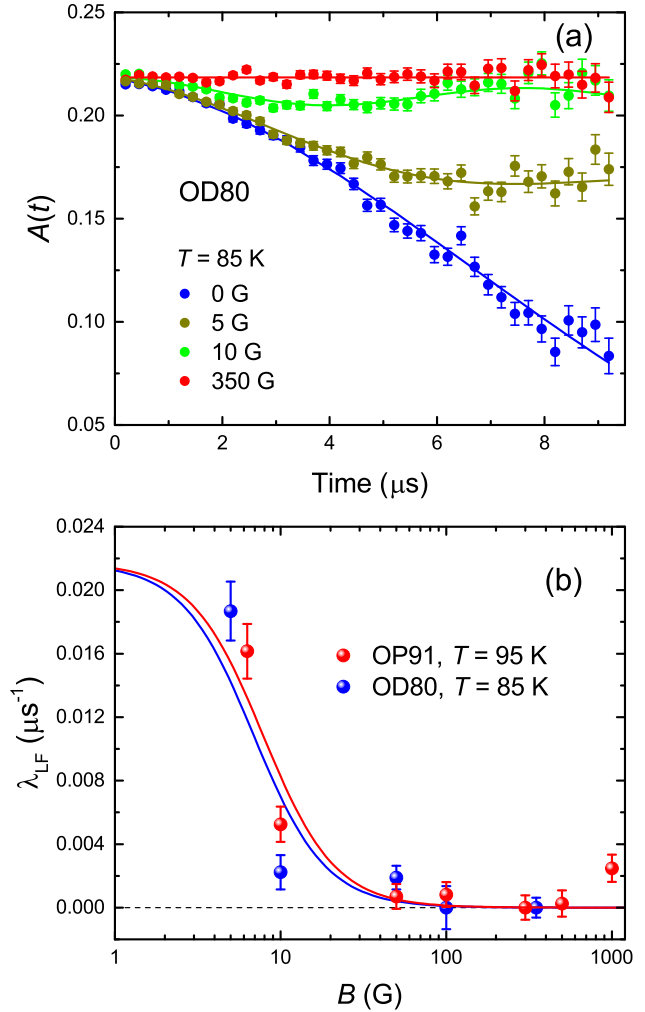}
\caption{(Color online) (a) Representative LF-$\mu$SR asymmetry spectra recorded on the OD80 sample at $T \! = \! 85$~K.
The curves are fits to Eq.~(\ref{eqn:LF}).
(b) Dependence of the LF exponential relaxation rate $\lambda_{\rm LF}$ on the magnitude of the applied magnetic field. 
The solid curves are fits to Eq.~(\ref{eqn:Redfield}), as described in the main text.
Note that the relaxation rates above 100~G are at the senstivity limit of $\mu$SR and may be compared with 
the values in Ag [inset of Fig~\ref{fig2}(a)].}
\label{fig3}
\end{figure}

To determine whether the local magnetic field detected in the PG region is static or fluctuating, we performed LF-$\mu$SR
measurements on each sample just above $T_c$. Figure~\ref{fig3}(a) shows LF-$\mu$SR asymmetry spectra for the OD80 sample 
with fits assuming the relaxation function
\begin{equation}
G_z(t) = G_{\rm KT}(B_{\rm LF}, \Delta, t) \exp(-\lambda_{\rm LF} t) \, .
\label{eqn:LF}
\end{equation}
Here $G_{\rm KT}(B_{\rm LF}, \Delta, t)$ is a static Gaussian LF Kubo-Toyabe relaxation function \cite{Uemura:99}.
The field dependence of the residual exponential relaxation rate $\lambda_{\rm LF}$ is shown 
in Fig.~\ref{fig3}(b). The data are well described by the Redfield theory expression \cite{Hayano:79}
\begin{equation}
\lambda_{\rm LF} = \frac{\gamma_\mu^2 \langle B_\mu^2 \rangle \tau}{1 + \left(\gamma_\mu B_{\rm LF} \tau \right)^2} \, ,
\label{eqn:Redfield}
\end{equation}
where $\gamma_\mu/2 \pi$ is the muon gyromagnetic ratio, $\langle B_\mu^2 \rangle$ is the mean of the square of the fluctuating
transverse field components, and $1/\tau$ is the characteristic fluctuation rate of the local fields $B_\mu$. Fits to this equation
yield $\langle B_\mu^2 \rangle^{1/2} \! = \! 1.4(2)$~G and $1/\tau \! = \! 0.7(1)$~MHz for the OP91 sample,
and $\langle B_\mu^2 \rangle^{1/2} \! = \! 1.3(3)$~G and $1/\tau \! = \! 0.6(2)$~MHz for the OD80 sample.
Thus the residual local internal magnetic field sensed by the muon is quasi-static and on the order of the resultant field of the
nuclear dipoles. In ortho-II Y123, NMR measurements place upper limits of 4~G and 0.3~G for any magnetic field fluctuating slower than 
$\sim \! 0.01$~MHz at the planar and apical oxygen sites, respectively \cite{Wu:15}.
A similar upper bound for static local fields at the apical oxygen site has been deduced from NMR on HgBa$_2$CuO$_{4 \! + \! \delta}$ 
(Hg1201) \cite{Mounce:13}. Hence, the weak quasi-static fields detected here in Bi2212 are likely fluctuating too fast to be detected by NMR.
 
Below 160~K where the implanted muon is immobile, the $T$-dependent $\lambda_{\rm ZF}$ may originate from a
continuous change in the nuclear dipole contribution or be caused by magnetic dipole moments of electronic origin.
The former may result from structural changes that modify the distance between the muon and nuclear spins, as
well as the direction of the maximal local electric field gradient (EFG) that defines the quantization axis for
the nuclear spins. A $T$-dependent electric quadrupolar interaction of the nuclei with the local EFG can also result 
from a gradual development of charge inhomogeneity or charge order. While $^{17}$O NMR 
measurements on overdoped Bi2212 ($T_c \! = \! 82$~K) demonstrate an inhomogeneous distribution of 
local EFG at the O(1) sites in the CuO$_2$ plane, this does not evolve with temperature \cite{Crocker:11}.
X-ray scattering measurements on underdoped Bi2212 show the development of CDW order within the 
PG phase \cite{Neto:14}, persisting as weak dynamic CDW correlations near $T^*$ \cite{Chaix:17}. 
Indeed, short-range CDW order has been identified in recent years to be ubiquitous in cuprates \cite{Comin:16}.
However, the CDW correlations are most pronounced in the underdoped regime and significantly weaken or
fade away near optimal doping. Moreover, in contrast to $\lambda_{\rm ZF}$ (Fig.~\ref{fig2})
the CDW correlations are suppressed below $T_c$. Hence CDW correlations do not seem to explain 
the $T$-dependent ZF relaxation rate observed below 160~K.

\begin{figure}
\centering
\includegraphics[width=8.5cm]{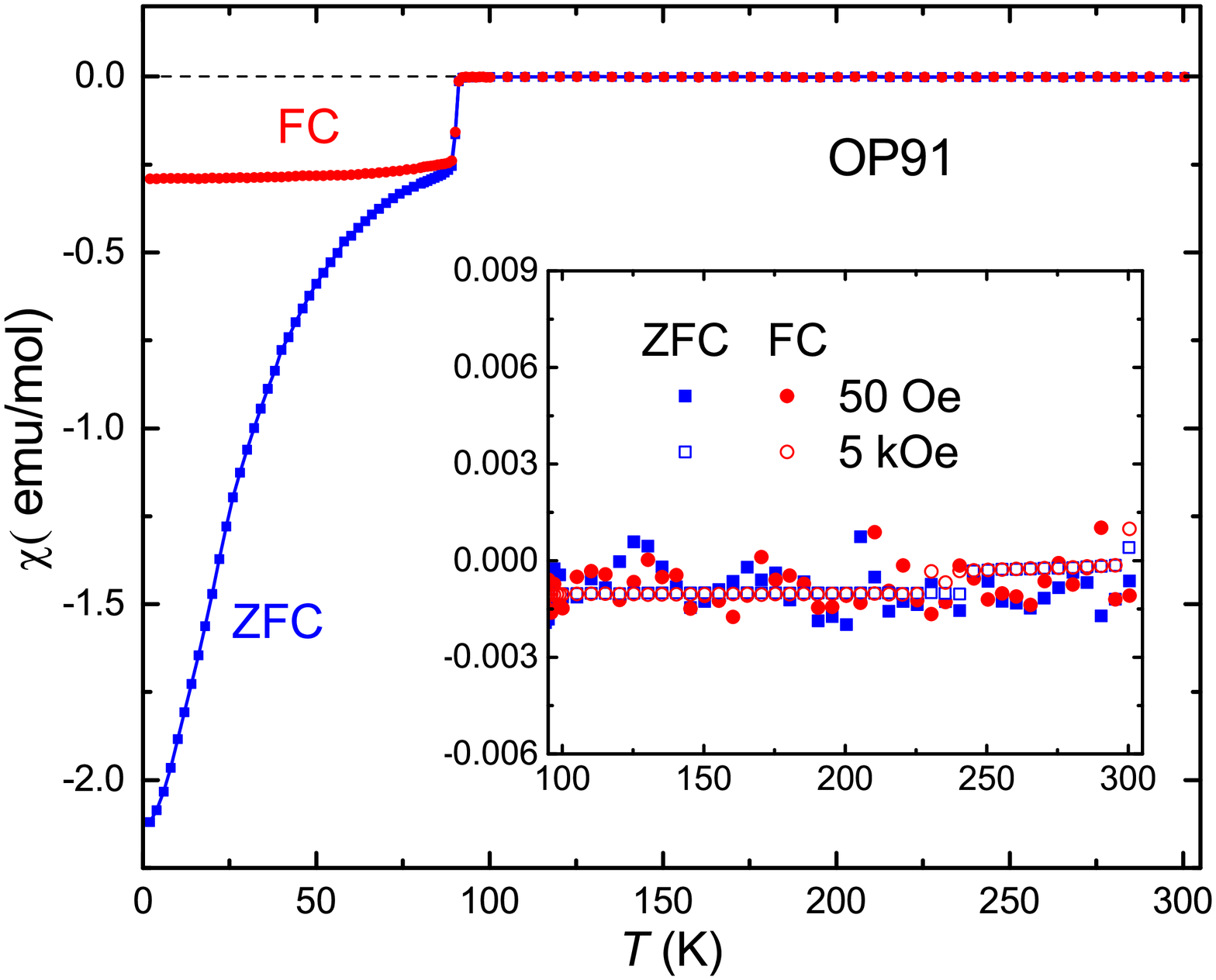}
\caption{(Color online) Temperature dependence of the bulk magnetic susceptibility for the OP91 sample
measured in an applied magnetic field of $H \! = \! 50$~Oe under field-cooled (FC) and zero-field cooled (ZFC)
conditions. The inset is a blow up of this data above $T_c$, and measurements taken at $H \! = \! 5$~kOe.}
\label{fig4}
\end{figure}

Another potential source of the ZF relaxation is dilute magnetic impurities. Dilute remnants of the underdoped 
phase containing Cu spin correlations fluctuating slow enough to be detectable on the $\mu$SR time scale are
unlikely to be present near and above optimal doping. Bulk magnetization measurements 
down to 2~K show no evidence of a magnetic impurity or secondary phase. As shown in Fig.~\ref{fig4}, the 
normal-state magnetic susceptibility of the OP91 sample measured up to 300~K exhibits no low-$T$ upturn indicative of 
trace amounts of a paramagnetic impurity. 

A previous weak LF-$\mu$SR study of overdoped polycrystalline samples of 
Bi$_2$Sr$_2$Ca$_{1-x}$Y$_x$Cu$_2$O$_{8 + \delta}$ in a LF of 20~G revealed a small 
increasing relaxation rate below $\! \sim \! 135$~K in a $T_c \! = \! 81$~K sample \cite{Tanabe:14}. While attributed
to an inhomogeneous distribution of internal magnetic field generated by
SC domains, the frequency scale for SC fluctuations above $T_c$ (10$^{10}$ to 10$^{14}$~Hz)
established by other methods \cite{Corson:99,Grbic:11,Bilbro:11} is too high to produce an observable
LF relaxation. Inserting $\langle B_\mu^2 \rangle^{1/2} \! \leq \! 20$~G and $1/\tau \! = \! 10^{10}$~Hz in
Eq.~(\ref{eqn:Redfield}) yields $\lambda_{\rm LF} \! \leq \! 3 \! \times \! 10^{-4}$~$\mu$s$^{-1}$, which is
far smaller that the LF relaxation rates reported in Ref.~\cite{Tanabe:14} and well below the reliable detection
limit. There is strong evidence for fluctuating SC domains in Bi2212 and Y123
well above $T_c$ from high transverse-field (TF) $\mu$SR experiments, which are highly sensitive to a distribution of 
time-averaged internal magnetic fields \cite{Lotfi:13}. Nonetheless, SC domains do not affect
the ZF relaxation rate.

The IUC magnetic order in Bi2212 inferred by neutrons is characterized by a pair of staggered magnetic
moments in the CuO$_2$ unit cell predominantly perpendicular to the CuO$_2$ plane and displaced from a Cu atom along
the [1, 1, 0] direction, with an ordered magnetic moment of $\sim \! 0.1$~$\mu_{\rm B}$ \cite{Thro:14}. 
While the precise {\it muon} site in Bi2212 is undetermined, the muon is expected to reside 
near an O atom. Calculations of the magnetic dipolar field generated by {\it static} IUC magnetic order 
at the {\it oxygen} sites in the CuO$_2$, SrO and BiO planes yield 3.1~G, 116~G and 1.4~G, respectively. Two of these 
values are on the order of the magnitude of the quasi-static internal field just above $T_c$ estimated from the LF-$\mu$SR data.
However, this does not exclude the possibility of magnetic order fluctuating too fast to be detectable on the 
$\mu$SR time scale, or equivalently $\gamma_{\mu}^2 \langle B_\mu^2 \rangle \tau \! \lesssim \!  0.001$~$\mu$s$^{-1}$. 
With this said, the detected field is close to the calculated size of the internal magnetic 
fields generated by {\it quasi-static} magnetoelectric quadrupolar ordering --- estimated to be $\sim \! 0.3$~G at the 
oxygen sites in Hg1201 \cite{Fechner:16} and potentially larger at the muon site(s) in Bi2212.

In summary, we have detected by $\mu$SR a weak $T$-dependent {\it quasi-static} internal magnetic field in 
the SC and PG phases of Bi2212, seemingly of electronic and intrinsic origin. While consistent with a 
static version of the IUC magnetic order inferred from neutrons measurements, this interpretation is difficult to 
reconcile with $\mu$SR studies of other cuprates. Our findings offer some support for a theory ascribing the 
primary order parameter in the PG phase to quasi-static magnetoelectric quadrupoles.   
 
\begin{acknowledgments}
We thank Marc-Henri Julien and Michael Fechner for insightful comments, and TRIUMF staff for technical assistance.
Sample preparation was performed at the CROSS user laboratories. The present work was partially supported by 
CIFAR, NSERC, and JSPS KAKENHI Grant Number JP15K17712.
\end{acknowledgments}

\end{document}